# Classification of Existing Virtualization Methods Used in Telecommunication Networks

Dmytro Ageyev, Oleg Bondarenko, Tamara Radivilova, Walla Alfroukh
Kharkiv National University of Radio Electronics, Nauky ave., 14, Kharkiv, Ukraine,
dmytro.aheiev@nure.ua, oleg.kh.bondarenko@gmail.com, tamara.radivilova@gmail.com,
http://nure.ua

*Abstract*—**This article studies the existing methods of virtualization of different resources. The positive and negative aspects of each of the methods are analyzed, the perspectivity of the approach is noted. It is also made an attempt to classify virtualization methods according to the application domain, which allows us to discover the method weaknesses which are needed to be optimized.**

## I. Introduction

Infocommunication industry is an important structural component of not only the economic and social life of the society, but also the process of informatization.

Investigating the statistics of mobile operators, it can be noted that the main role in the network now is not by voice traffic, but by Internet traffic or data traffic [1].

The growing amount of providing infocommunication services, the increased flexibility of their implementation and the increasing volume of transferred data require constant development and improvement of infocommunication infrastructure. Accessibility, security, fault tolerance and the ability to scale on-demand services have become critical indicators of modern infocommunication infrastructures.

Another important point is the cost of services, which directly depends on the cost of equipment and the efficiency of its use. Minimizing the cost of services for the end user allows you to capture as much of the market as possible and attract more customers.

To ensure the enumerated requirements for infocommunication infrastructure, hardware and software virtualization technologies are often used [2]. These technologies have proven themselves to solve problems associated with increasing the utilization of physical computing or network resources, as well as with optimal management of the network topology or application infrastructure. Virtualization technology is widely used in the concepts of cloud computing, multiservice SDN networks, private VPN, VLAN, VXLAN networks [3, 4].

The improving of infocommunication infrastructure requires the development of optimal management methods and methods for the synthesis of infocommunication systems. The upgrading of previously known methods and developing new highly efficient methods of control and synthesis require the classifying of virtualization methods with identification of common features and differences, choosing approaches and concepts for building of virtualized infocommunication systems.

In this article we analyzed the main principles of virtualization, proposed their classification, highlighted the advantages and disadvantages, as well as the features of each method.

## II. The Origin of the Term

One of the concepts is to start one or more guest operating systems on the host. However, if we consider the issue a little deeper, then this definition is too specific. There is a large number of services, hardware and software, that can be put on work upon using virtualization technologies.

In fact, virtualization is used to abstract real physical resources from the end user, i.e. hiding real physical indicators of equipment. This can be a representation of one physical resource (server, network resource, data store, RAM, operating system) in the form of several virtual resources, or a representation of several physical resources in the form of a single virtual resource.

IBM started using virtualization in its research systems to organize multiple access. The concept of a virtual machine provided the separation of user space by simulating a standalone computer for each person. In the early 90's, with the advent of microprocessor technologies on the market, the concept of single large mainframes began to recede into the background and x86/x64 standard workstations appeared. As a result, virtualization, as the idea of sharing access to large resources, has not been using until the end of the 90s.

## III. Types of Virtualization

Virtualization is divided into two main types. The first type is platform virtualization, which is responsible for the virtualization of processes and applications. The second type is virtualization of resources, which involves splitting or combining real physical resources into virtual groups or nodes.

### A. Platform Virtualization

Platform or server virtualization is the most common and claimed type of virtualization that is used by large corporations and companies that provide a cloud computing resource. In case of server virtualization, one physical machine is divided into multiple virtual environments. While solving the task of platform virtualization, the following approaches can be distinguished:
- Hypervisor usage;
- Emulation;
- Containerization.

The hypervisor [5] is software that intercepts the system calls of the virtual machine and allows the virtual environment to interact with real physical equipment.

Hypervisors are divided into several types depending on the kind of virtualization:

Type 1 – this type of hypervisor is installed directly on the server hardware, fig. 1. The virtual resource sets are created and the main virtual machine is installed - the master that is designed to monitor the status of all other virtual machines. Virtual machines work directly with the server hardware through a hypervisor. Thanks to this method, the capacity gain up to 10% compared with the next option [6].

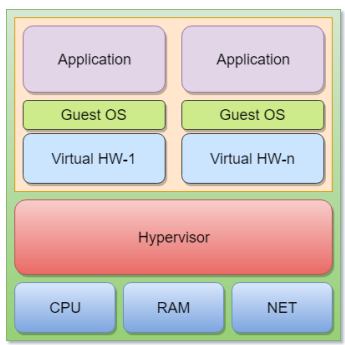

Figure 1. First type of hypervisor-constructed machine

First type of hypervisor-constructed machine is used in paravirtualization and hardware virtualization.

**Paravirtualization** uses a first type of hypervisor that is installed on bare metal, at the same time a simplified image of the operating system is installed. The last one is used to control and monitor the other virtual machines. This kind of virtualization is not supported by all available platforms, as it requires modification of the kernel of the guest operating system. When the system is modified, an additional set of APIs is included in the kernel. It can work directly through it with the hardware of the server, without conflicting with other virtual machines. In this case, there is no need to use the host operating system as an intermediary. As a result, the charges are reduced, and the virtual machine can use the physical resource more effectively.

**Hardware virtualization** is the evolution of the software platforms` abstraction levels. It's about moving from multitasking techniques and dividing physical resources over time, to multithreading using multiple cores of a physical processor. Next after multithreading is the level of hardware virtualization, the method of dividing physical resources between processes. Multitasking is the first level of application abstraction. Each application shares the resources of the physical processor in the mode of dividing the execution of the code in time. Multithreading technology is a hardware virtualization technology and allows to distribute the load caused by the process in the system between the real physical cores of the processor. Hardware virtualization is a logical extension of the two previous methods and allows to virtually split the processes of different operating systems between the physical cores of the CPU.

Type 2 - this type of hypervisor runs on top of the native operating system that is installed on the server, fig. 2. The main task of the hypervisor in this case is to link the operating systems of the virtual machines to the host operating system in the dynamic broadcasting mode. In this mode, approximately 15-20% of the capacity goes to the coordination between the operating systems [6].

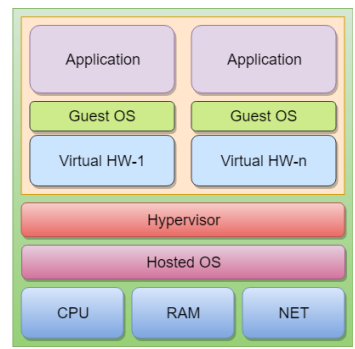

Figure 2. Second type of hypervisor-constructed machine

**Full virtualization (Dynamic translation)** is one of the virtualization kinds is based on second type of hypervisor. Hypervisor is installed in the host operating system. After that, sets of virtual physical resources and virtual machines are created. Using built-in operating systems on a virtual (guest) machine. A virtual machine works with real physical hardware through a hypervisor and host operating system.

**Emulation.** Is used to develop outdated or hard-to-access platforms software. In this case, the hardware part of the required platform is fully reproduced by software for one of the existing operating systems (most often - x86 compatible). The work of the processor and registers, all peripheral devices, drivers is emulated. All management functions are assumed by the parent operating system.

**Host-level virtualization (Containerization).** This approach provides using one operational system kernel to create several independent parallel systems. The principle is alike applications in the sandbox, though in a limited space not an application is launched, but an operating system that can work with the kernel of the parental system. For guest software, only private network and hardware environments are created, fig. 3 [2].

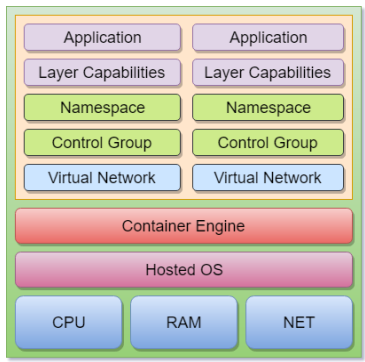

Figure 3. Host-level virtualization

Platform Virtualization, undoubtedly, has a lot of advantages:
- System deployment speed. The process of creating a virtual machine takes from one to 15 minutes, depending on the required configuration and the initial settings of the system.
- Compatibility. The same image of the operating system can be used to create virtual machines on different physical servers.
- Increase in the percentage of equipment utilization - this results in saving money on hardware, reducing administrative costs and saving energy.
- Security. To restore compromised virtual machines, snapshots or clean images of operating systems are used. The disk space in which the virtual machine runs is isolated from the real one.
- Development. Virtual machines are often used in the software development, because they allow you to reproduce the necessary environment quickly, which consists of a set of virtual machines, and test any changes.

Server virtualization has several disadvantages:
- Security. The benefits of virtualization are also its disadvantages, as additional vulnerability points are created. The compromised image of the operating system can be easily distributed over a virtual network channel or with the help of discovered vulnerabilities in the hypervisor.
- Administration. Because multiple virtual machines can run on the same physical server, you may need skills to automate and administrate a large number of machines.
- Licensing. Some of the operating systems do not support the distribution of the activation keys to virtual machines, therefore, it may be necessary to purchase additional licenses.
- Performance. The program layer in the form of a hypervisor needs to use a small part of the computing resource of a real server. You should always consider the performance margin by 5-20%, depending on the type of a virtualization used.

*B. Virtualization of resources*

**Network virtualization** is a very popular form of virtualization. This concept implies the integration of hardware and software network resources into a single network complex or a set of network functions and objects. This type of virtualization is considered to be external and can be used for virtual integration of remote physical objects into a single network topology.

Virtualization in this case can be as network devices - routers, switches and multiplexers, and network interfaces. As well similar set of virtualized network resources allows you to create virtual private networks (VPNs) and virtual local area networks (VLANs) [7].

VPN is used to securely connect remote clients to a work or home network, to simulate a direct connection and to join several corporate networks into one domain zone.

VLAN is used to distinguish traffic within the enterprise LAN. On the physical network, virtual channels are created, which are followed by packets of different user groups.

Among the many advantages of network virtualization are the following:
- Access settings. The need to build their own physical communication channels between branches of one company has disappeared, administrators can quickly create virtual networks and configure access parameters such as routing, bandwidth control and quality of service.
- Consolidation. Physical networks can be combined into one virtual network for shared access and management.

Like virtualization of servers, network virtualization can make the topology of the network infrastructure more complicated, introduce some performance overhead and require special skills from system administrators.

**Storage virtualization** is a very common and most popular type of virtualization. The process of storage virtualization involves abstracting user data from an actual physical information storage [8].

Rather popular technology RAID provides this functionality only at a basic level, multiplying files on several physical media, which are combined into logical massive.

The term "storage virtualization" is a broader concept and in addition to backing up data, it includes migration, caching and secure data transport capabilities at the customer's request.

Virtual storage can be classified by the type of data synchronization. In terms of implementation, each service provider has its own approach, but in general, synchronization can be symmetrical and asymmetric.

Taking in account the specifics and complexity of this type of virtualization, we can distinguish the following advantages:
- Migration. Thanks to a large set of technologies, data can easily be transferred from one storage to another, even if physically they are on different continents.
- Scalability and management. With the change in the number of data accesses, the number of virtual storages can be automatically increased or decreased.
- Reservation. A well-tuned set of logical stores allows you always to have a backup copy of the data.

Disadvantages of storage virtualization:
- Metadata. As one virtual storage can reside on multiple physical media, you must store the service files to match the disk layouts.
- Lack of standards and compatibility. Due to the fact that storage virtualization is a concept, not a standard for a large number of vendors, there is no interaction, so you need to consider this fact when designing.

**Application virtualization** is in high demand last time. Depending on the initial requirements, different levels of application detail are distinguished. Virtually it can be deployed as an entire infrastructure of interacting applications, and a small part of the application is implemented as a separate function whose task is to process requests and produce results depending on the embedded logic.

Virtualization of the application makes the software independent from the physical operating system and hardware. This simplifies the process of scaling and deploying the application. Virtualization of the application makes it possible to run several incompatible applications simultaneously on the same computer, or rather in the same operating system or a set of operating systems [9].

This type of virtualization is a set of the above types of virtualization.

Advantages of this type of virtualization:
- Cross-platform applications. To run the application, the system administrator does not need to check the application for compatibility; you only need to take care of allocating enough computing resources.
- Speed of installation. Since there is no need to ensure compatibility between multiple operating systems, the installation process takes a few seconds.

IV. CONCLUSIONS

The methods of virtualization used in infocommunication systems are divided into two types: virtualization of platforms and virtualization of resources, which in turn are divided by kinds. For virtualization platforms we defined: emulation, dynamic translation, paravirtualization, hardware virtualization and containerization; for virtualization of resources: network virtualization, storage virtualization and application infrastructure virtualization.

The advantages of all types of virtualization include the higher efficiency of using limited system resources due to its dynamic distribution between processes, as well as higher security of data and processes due to their isolation at a lower level than the processes themselves.

All methods of virtualization are characterized by the principles of consolidation or separation of real physical resources for the end user. To implement this approach, specialized software is used. It performs the role of manager and monitors the state of the equipment. This leads to the complexity of the system, requires additional system resources, which can be noticed as the disadvantages of virtualization. In turn, the task of allocating physical resources between virtual objects appears, when the manager is working. It requires the usage of optimization methods to improve the effectiveness of virtualization methods.

The development of methods for optimal control and synthesis of virtualized infocommunication systems requires a preliminary solution of problems: an assessment of the applicability of existing optimization methods in virtualization environments, as well as the choice of a mathematical apparatus for solving optimization problems.